\documentclass{webofc}
\usepackage[varg]{txfonts}   
\woctitle{?????????}
\usepackage{verbatim} 
\usepackage{epstopdf}

\usepackage{lineno,hyperref}

\usepackage{glossaries} 
\makeglossaries
\newacronym{UHECR}{UHECR}{ultra high energy cosmic rays}
\newacronym{EAS}{EAS}{extensive air shower}
\newacronym{sd}{SD}{surface detectors}
\newacronym{fd}{FD}{fluorescence detectors}
\newacronym{MPD}{MPD}{muon production depth}

\newcommand{\epos}{EPOS-LHC\xspace}
\newcommand{\qgs}{QGSJ{\textsc{et}}II-04\xspace}

\begin{document}
\title{Constraints on hadronic models in extensive air showers with the Pierre Auger Observatory}

\author{\firstname{Jo\~{a}o} \lastname{Espadanal}\inst{1}\fnsep\thanks{\email{jespada@lip.pt } } for the Pierre Auger Collaboration\inst{2}}

\institute{LIP, Av. Elias Garcia 14-1, 1000-149 Lisboa, Portugal
\and
Pierre Auger Observatory, Malargue, Argentina {\scriptsize(Full author list: \url{http://www.auger.org/archive/authors\_2016\_08.html})}}

\abstract{
Extensive air showers initiated by ultra-high energy cosmic rays are sensitive to the details of hadronic interactions models, so we present the main results obtained using the data of the Pierre Auger Observatory. The depth at which the maximum of the electromagnetic development takes place is the most sensitive parameter to infer the nature of the cosmic rays. However, the hadronic models cannot describe consistently the maximum and the muon measurements at energies higher than those reached at the LHC.
}
\maketitle
%
%
\section{Introduction}
\label{intro}%
Particle physics at energies beyond the reach of accelerators can be studied with air showers induced by \gls{UHECR}. Data from the Pierre Auger Observatory \cite{PAO,PAOold}, indicate that the hadronic interaction models \epos and \qgs have a muon deficit of around 30\%  to 80\% \cite{MuonN2014}, respectively, to match the primary composition inferred from the depth of the electromagnetic (EM) shower maximum.
%
A \gls{UHECR} collides with an atmospheric nucleus and initiates an \gls{EAS} of secondary particles. After the first interaction $\sim80\%$ of the particles produced are pions and $\sim8\%$ are kaons. In each hadronic interaction, the produced $\pi^0$ carry around $25\%$ of the energy, and they immediately decay into photons, feeding the electromagnetic cascade. The rest of the energy is carried mainly by charged pions (and also other mesons and baryons), which continue the hadronic cascade. The mesons propagate until their energy drops below their critical energy, $ \mathcal{O}$(100 GeV), and then they decay into muons among other things.
The electromagnetic development and muonic content depend on the primary composition and the hadronic interaction properties, such as the fraction of EM energy per interaction, multiplicities, hadronic cross sections, etc. 
So, assuming that the composition can be properly understood, a discrepancy between observed and predicted muon numbers, must indicate an inaccurate description of the hadronic models.
 
 The Pierre Auger Observatory is the largest \textit{hybrid} cosmic-ray detector. It has two main detection techniques: the \gls{sd}
 which consists of an array of 1660 water-Cherenkov stations extended over $\sim3000\text{ km}^2$
  with an almost 100\% duty cycle, and the \gls{fd} composed of 27 telescopes operating with a $\sim15\%$ duty cycle. To understand the muon puzzle, an upgrade is planned for the Observatory\cite{AugerPrime}, to be able to separate the muonic and electromagnetic components of the shower at the ground. Due to the limited space available here, we will briefly discuss the main features of muon studies and electromagnetic profiles.
%
%
%
\section{The EM cascade}%
\label{sec: The EM cascade}
The longitudinal development of the energy deposit in air by the showers can be measured in the \gls{fd}, as a function of the atmospheric slant depth. The profile is dominated by the electrons belonging to the EM cascade and the profile maximum is denoted by \textit{$X_{max}$}. The depth of shower maximum is mostly sensitive to the first interaction position and the primary particle type, i.e. showers induced by heavy primaries will develop higher (at a shallower depth), faster and with less shower-to-shower fluctuations than those induced by lighter nuclei (due to a larger interaction cross-section and higher number of nucleons).
 The primary composition can be inferred statistically from the distribution of shower maxima, due to the fluctuations in the first few hadronic interactions in the cascade. The $X_{max}$ distributions can be a superposition of different nuclei, so four component (proton, iron, helium and nitrogen) were considered to fit the data distributions. For both models considered, \qgs and \epos, the fraction of protons is decreasing and the fraction of helium is increasing for energies above a few EeV, with no significant fraction of iron for all energies (see \cite{XmaxInt2014}).
 
\begin{figure}[h]
\centering
\includegraphics[width=0.405\linewidth,clip]{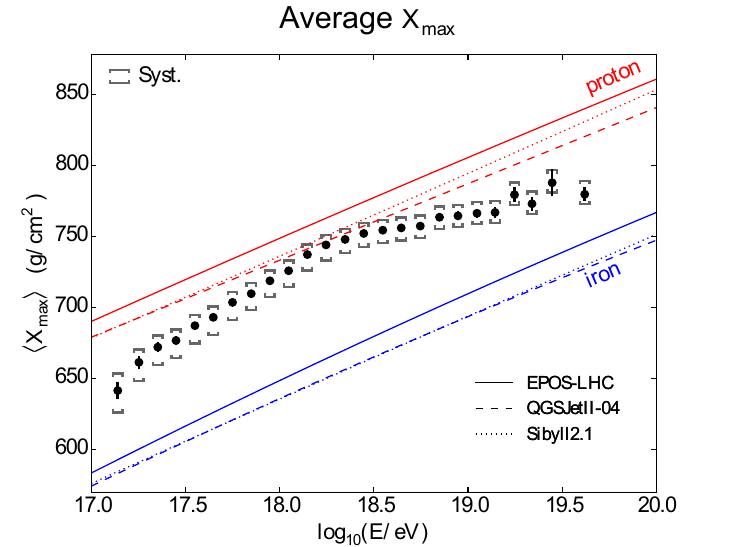}
\includegraphics[width=0.405\linewidth,clip]{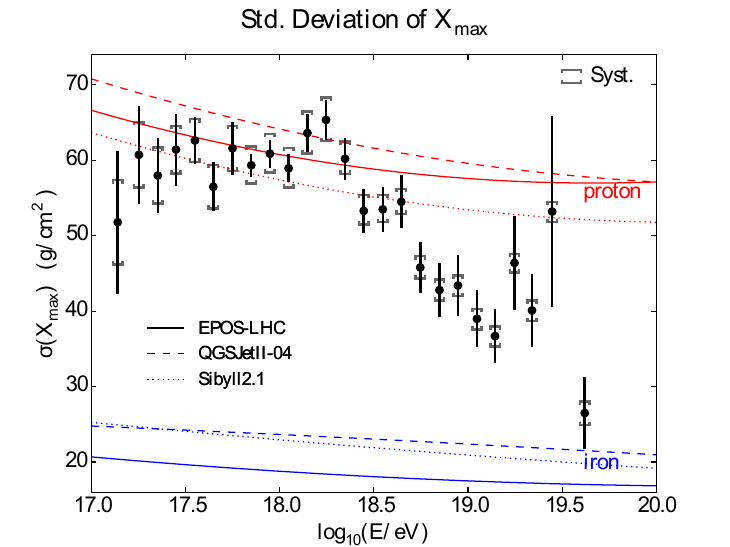}
\caption{The mean (left) and the standard deviation (right) of the measured $X_{max}$ distributions as a function of energy compared to air-shower simulations for proton and iron primaries \cite{Xmax2014,ICRC2015}.}
\label{fig: xmax}      
\end{figure}

In figure \ref{fig: xmax} left, the energy evolution of the average $\langle X_{max}\rangle$ is shown for the data and predictions of the hadronic interaction models. Different selection cuts are applied to ensure good data-taking conditions\cite{Xmax2014}.
The elongation rate ($D_{10}=\text{d}\langle X_{max}\rangle/\text{d}\log_{10}E$) would be constant if the composition were pure. The best fit to our data comprises two linear fits (in units of $\log_{10}(E/\text{eV})$) with a breaking point at $\log_{10}(E/\text{eV}) = 18.27\pm 04(stat)^{+0.06}_{-0.07}(sys)$.
Considering the model predictions, above the  breaking point, the observed rate of change of $\langle X_{max}\rangle$ becomes significantly smaller ($\sim 26 \text{ g.cm}^{-2}/decade$) indicating that the composition is becoming heavier.
The observed width of the $X_{max}$ distribution is corrected by subtracting the detector resolution in quadrature to obtain $\sigma(X_{max})$  (figure \ref{fig: xmax} right).
It also shows that above the breaking point the composition changes from light elements to heavier ones.
%
%
%
\vspace{-0.35cm}
\section{The hadronic cascade}%
\begin{figure}[!b]
\centering
\includegraphics[width=0.37\linewidth,clip]{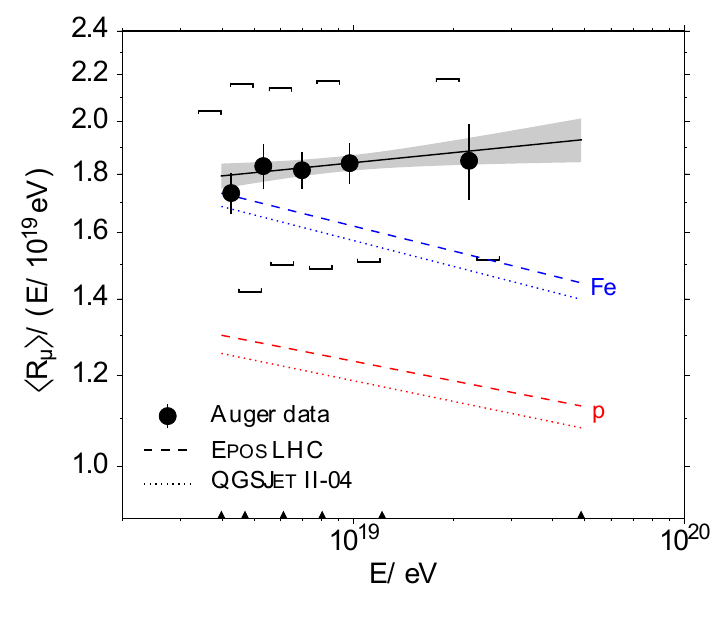}\hspace{0.05\linewidth}
\includegraphics[width=0.39\linewidth,clip]{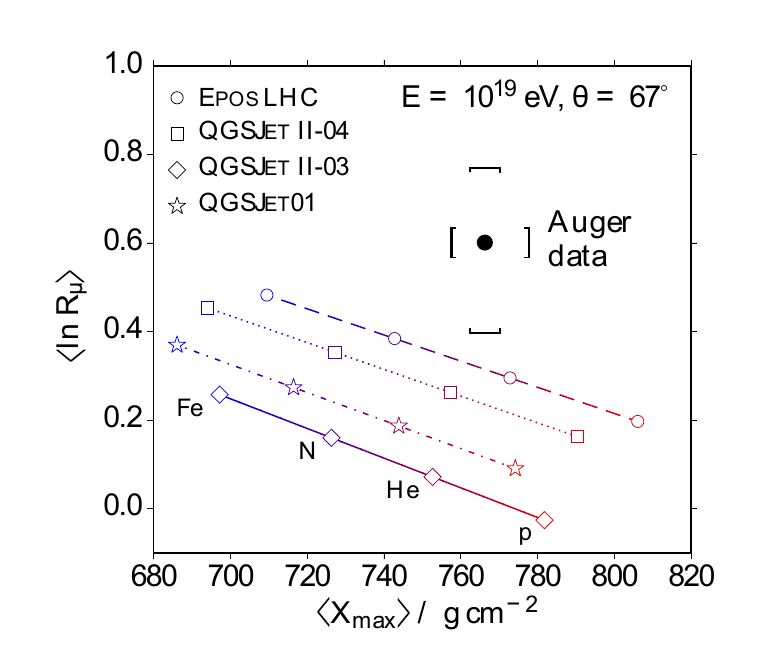}
\caption{(Left) $\langle R_\mu \rangle$ vs. primary energy, compared to air-shower simulations. (Right) $\langle \ln R_\mu \rangle$ vs. $\langle X_{max} \rangle$. Representative primary masses are indicated by open symbols \cite{MuonN2014,ICRC2015}.}
\label{fig: MuonN}
\vspace{-0.25cm}
\end{figure}

\label{sec: The hadronic cascade}
In highly-inclined events, the electromagnetic component of a shower is largely absorbed in the atmosphere, while the muons still survive to the ground, which is suitable for direct studies of the EAS muon content. Hybrid events have independent energy and muon measurements, and with zenith angles in the range $62^\circ< \theta < 80^\circ$, the SD signals can be considered as muons.
The whole procedure can be seen in \cite{MuonN2014}, schematically: for each event, the muon density at the ground $\rho_\mu^{rec}$ is recorded with the SD and adjusted to the shape $\rho_\mu^{map}$ and relative muon content $R_\mu$, as $\rho_\mu^{rec}=R_\mu\;\rho_\mu^{map}$. The footprint $\rho_\mu^{map}$ is derived from simulations of protons with QGSJ{\textsc{et}}II-03 at an energy of $E = 10^{19}$ eV.
A power-law dependence of the mean muon content with energy is expected as $\langle R_\mu \rangle =a (E/10^{19}\text{eV})^b$. The data shown in figure \ref{fig: MuonN} left, gives $a = 1.84 \pm 0.03(stat) \pm 0.32(sys)$ and
$b = 1.03 \pm 0.02(stat) \pm 0.03(sys)$. Even with the large systematic uncertainty (square brackets) data are not compatible with a pure proton composition, and, at higher energy, are marginally compatible with iron. The energy evolution slope also suggests a transition from a lighter to a heavier composition. 
\begin{figure}[!b]
\centering
\includegraphics[width=0.36\linewidth,clip]{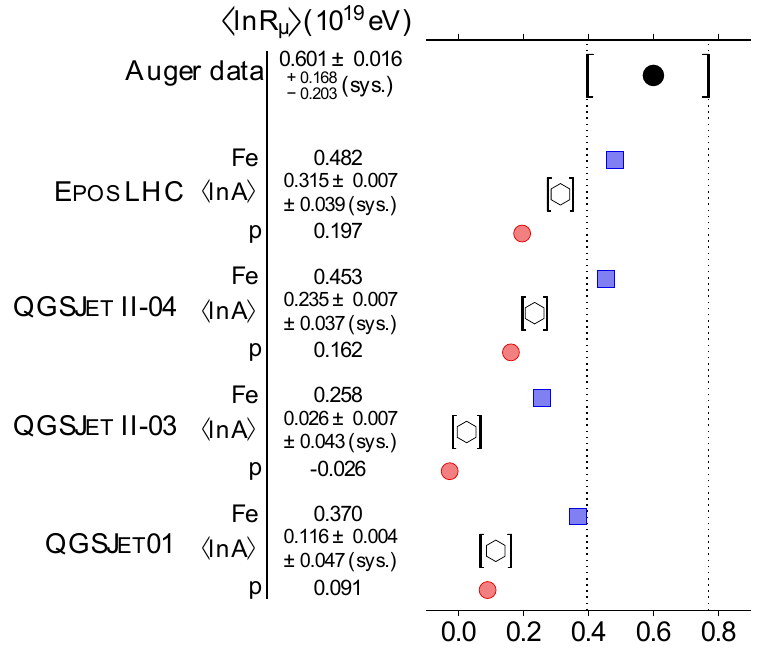}\hspace{0.05\linewidth}
\includegraphics[width=0.36\linewidth,clip]{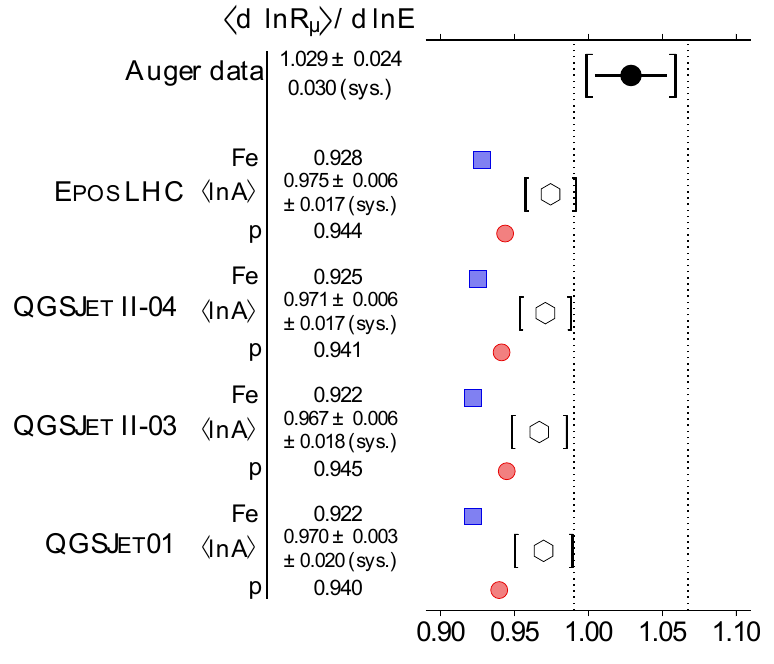}
\caption{Comparison of $\langle \ln R_\mu \rangle$ (left) and $\frac{d\langle\ln R_\mu \rangle}{d\ln E}$ (right) between $4\times 10^{18}$ eV and $5\times 10^{19}$ eV with predictions for air-shower simulation models for a pure proton, a pure iron and a mixed composition compatible with the FD measurements (labeled as $\langle \ln A \rangle$) \cite{MuonN2014,ICRC2015}.}
\label{fig: AvMouns}
\end{figure}

The problems with the hadronic models becomes evident when the muon results are compared to the depth of shower maximum (electromagnetic). In figure \ref{fig: MuonN} right, the average $\langle X_{max}\rangle$ and $\langle R_\mu\rangle$ are shown for the data and simulations at $E=10^{19}$ eV, where the Auger data fall completely out of the model's phase-space.
The average values of $\langle R_\mu \rangle$ and the logarithmic slope $\frac{d\langle\ln R_\mu \rangle}{d\ln E}$ are shown in figure \ref{fig: AvMouns} and compared to the models predictions using the EM $X_{max}$ composition. 
In all cases the models fail to match both measurements, requiring a substantial increase in muon production. Additionally, the large value of the measured slope favors a changing composition.

\begin{figure}[h]
\centering
\includegraphics[width=0.40 \linewidth,clip]{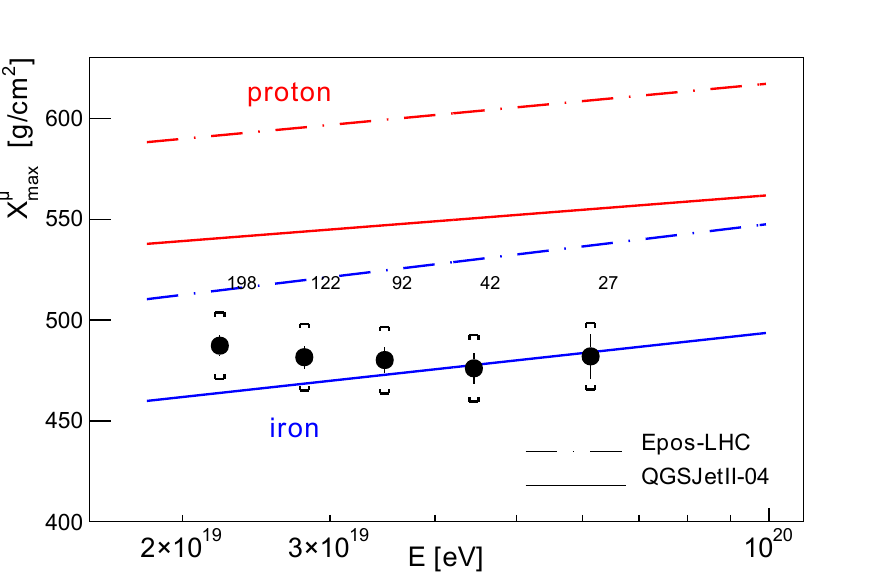}
\includegraphics[width=0.27\linewidth,clip]{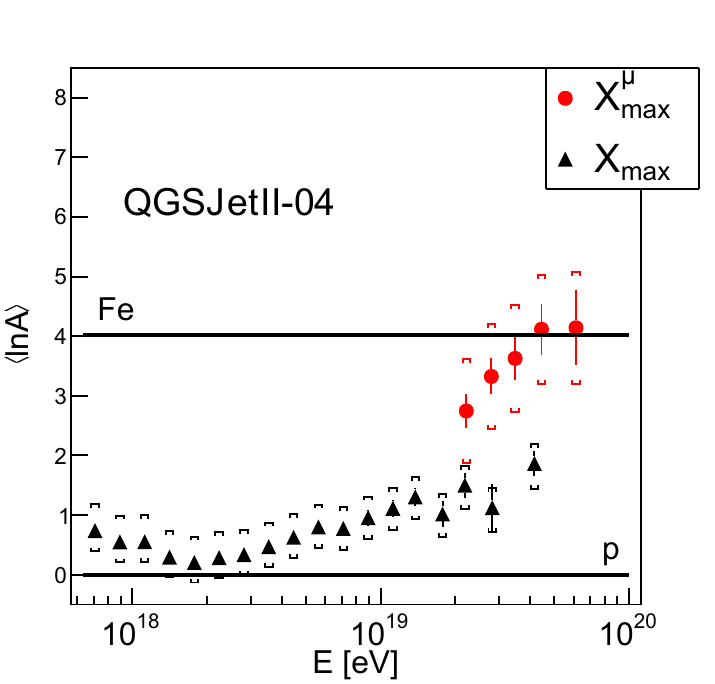}
\includegraphics[width=0.27\linewidth,clip]{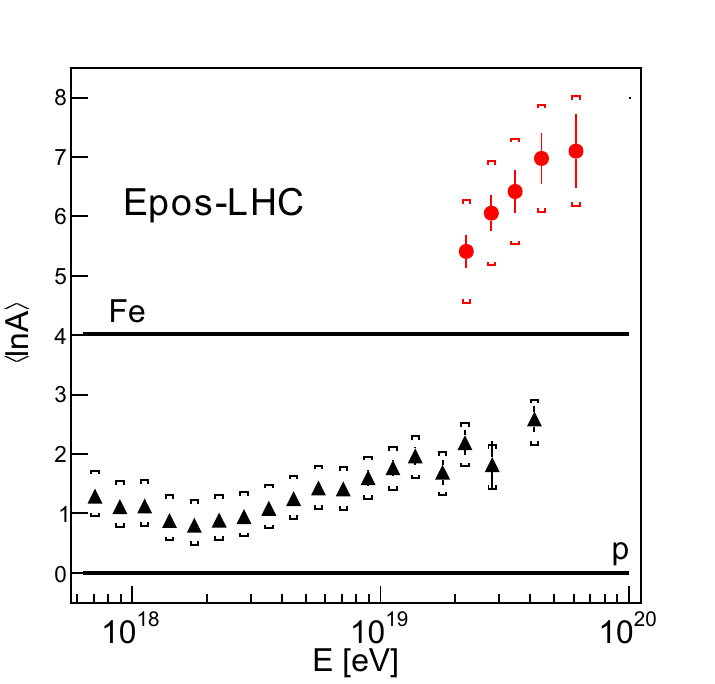}
\caption{Evolution of $X^{\mu}_{max}$ with energy\cite{MPD2014}(left), the number of events is indicated in each energy bin and brackets representing the systematic uncertainty. $\langle \ln A \rangle$ from $\langle X^{\mu}_{max} \rangle$ and $X^{\mu}_{max}$ as a function of energy\cite{MPD2014,ICRC2015}(right).}
\label{fig: Xmumax} 
\vspace{-0.5cm}
\end{figure}

\label{subsec: Muon production depth}
The EM component is highly absorbed in the atmosphere, so under certain conditions (large zenith angles, large distances to the shower core), the SD signal is dominated by muons. In those cases, the temporal structure and distance to the shower axis of muons arriving at the ground can be measured. Using a set of simple assumptions\cite{MPD2014}, they can be used to obtain the production point of the muons along the shower axis. 
Summing all muons, the longitudinal development of \gls{MPD} can be measured on an event-by-event basis. The evolution with energy of the corresponding maximum $X_{max}^\mu$ is shown in figure \ref{fig: Xmumax} left. 
Both muonic $\langle X_{max}^\mu\rangle$ and electromagnetic $\langle X_{max}\rangle$ parameters can be translated into $\langle \ln A \rangle$ as shown in figure \ref{fig: Xmumax} right, for two interaction models, allowing a direct comparison of both observables. The \qgs model and the \epos model give incompatible mass values, from the electromagnetic and muonic components.

%
%
%
%
%
\section{Conclusions}%
\label{sec: Conclusions}
Using data from the Auger Observatory we have found that the average primary composition from the $X_{max}$ observable is not consistent with a pure composition, assuming the overall correctness of the hadronic interaction models considered. Under the assumption that there is no new physics affecting the air-shower development trans-LHC energies, we can see a transition from light to heavy UHECR composition. 
Nevertheless, muon measurements are inconsistent with hadronic interaction
model predictions. The models yield a deficit in the produced number of muons and the interpretation of the longitudinal maxima $\langle X_{max}^\mu\rangle$ and $\langle X_{max}\rangle$ leads to an inconsistent average primary mass $\langle \ln A \rangle$. Neither electromagnetic or muonic measurements can be described consistently by the hadronic interaction models, so they can be used to constrain them at energies beyond existing terrestrial accelerators.

\subsection*{Acknowledgments}
Thanks to LIP and FCT for a Postdoc grant; this work would not have been possible without the strong commitment and effort of the technical and administrative staff in Malarg\"ue.

\bibliography{myref}

\begin{thebibliography}{8}

\bibitem{PAO}
A.~Aab, et~al. (Pierre Auger Collaboration), Nucl. Instrum. Meth.
  \textbf{A798}, 172 (2015), \texttt{1502.01323}

\bibitem{PAOold}
J.~Abraham, et~al. (Pierre Auger Collaboration), Nucl. Instrum. Meth.
  \textbf{A523}, 50  (2004)

\bibitem{MuonN2014}
A.~Aab, et~al. (Pierre Auger Collaboration), Phys. Rev. \textbf{D91}, 032003
  (2015)

\bibitem{AugerPrime}
A.~Aab, et~al. (Pierre Auger Collaboration), arXiv  (2016), \texttt{1604.03637}

\bibitem{XmaxInt2014}
A.~Aab, et~al. (Pierre Auger Collaboration), Phys. Rev. \textbf{D90}, 122006
  (2014), \texttt{1409.5083}

\bibitem{Xmax2014}
A.~Aab, et~al. (Pierre Auger Collaboration), Phys. Rev. \textbf{D90}, 122005
  (2014), \texttt{1409.4809}

\bibitem{ICRC2015}
A.~Aab, et~al. (Pierre Auger Collaboration), Proc. of the 34th ICRC  (2015),
  \texttt{1509.03732}

\bibitem{MPD2014}
A.~Aab, et~al. (Pierre Auger Collaboration), Phys. Rev. \textbf{D90}, 012012
  (2014), \texttt{1407.5919}

\end{thebibliography}
%

\end{document}